\documentclass[journal]{IEEEtran}
\usepackage[symbol]{footmisc}
\usepackage{graphicx}
\usepackage{xcolor}
\usepackage{tabularray}
\usepackage{mdframed}
\usepackage{amsfonts}
\usepackage{amsmath}
\usepackage{algorithm}
\usepackage{algorithmicx}
\usepackage{algpseudocode}
\usepackage{acronym}
\usepackage{balance}
\usepackage{pbalance}
\usepackage{longtable}
\usepackage{hyperref}
\usepackage{amssymb}
\usepackage{pifont}
\definecolor{darkgreen}{rgb}{0.0, 0.5, 0.0}
\newcommand{\cmark}{\textcolor{darkgreen}{\ding{51}}}%
\newcommand{\xmark}{\textcolor{red}{\ding{55}}}%

\usepackage[switch]{lineno}
\usepackage{svg}
\newcommand{\TODO}[1]{\textcolor{red}{#1}\GenericWarning{}{LaTeX Warning: TODO: #1}}\newcommand\todo\TODO

\mdfdefinestyle{theoremstyle}{%
linecolor=blue!50,linewidth=1.5pt,%
frametitlerule=true,%
frametitlebackgroundcolor=gray!20,
innertopmargin=\topskip,
nobreak=true,
}

\newenvironment{RQbox}[1]{%
  \begin{mdframed}[style=theoremstyle,frametitle={#1},linecolor=lightgray]
}{%
  \end{mdframed}
}

\title{On-Chain Analysis of Smart Contract\\ Dependency Risks on Ethereum}

\author{Monica Jin$^\dagger$, Raphina Liu$^\dagger$ and Martin Monperrus\\KTH Royal Institute of Technology\\
\texttt{\{mjin, raphina, monperrus\}@kth.se}}
\date{June 2024}

\begin{document}

\maketitle
\begin{abstract}
In this paper, we present the first large-scale empirical study of smart contract dependencies by analyzing over 41 million contracts and 11 billion interactions on Ethereum up to December 2024.
Our results yield four key insights: 
(1) 59\% of contract transactions involve multiple contracts (median of 4 per transaction in 2024), indicating potential smart contract dependency risks;
(2) the ecosystem exhibits extreme centralization, with just 11 (0.001\%) deployers controlling 20.5 million (50\%) of alive contracts, with major risks related to factory contracts and deployer privileges; 
(3) three most depended-upon contracts are mutable, meaning large parts of the ecosystem rely on contracts that can be altered at any time, which is a significant risk;
(4) actual smart contract protocol dependencies are significantly more complex than officially documented, undermining Ethereum's transparency ethos, and creating unnecessary attack surface.
Our work provides the first large-scale empirical foundation for understanding smart contract dependency risks, offering crucial insights for developers, users, and security researchers in the blockchain space.

\end{abstract}

\footnotetext{$^\dagger$ Equal contribution to the work.}

\section{Introduction}
\label{sec:intro}

Blockchain~\cite{nakamoto2008bitcoin} is a decentralized digital ledger that records transactions transparently.
It enables verifiable interactions between smart contracts through transactions.
Smart contracts~\cite{ethereumIntroductionSmart} are programs encoding self-executing agreements with the terms directly written into code.
Ethereum~\cite{ethereumEthereumorgComplete} is the most popular smart contract platform.
It has been designed with the vision that smart contracts interact with each other, as part of ever more complex and composable agreements \cite{wood2014ethereum}.

Understanding how smart contracts depend on each other is crucial yet understudied.
In traditional software ecosystems, tools like package managers, software bill of materials, and automated dependency bots are used to track and manage dependencies.
However, these methods do not apply to smart contracts.
The smart contract ecosystem lacks standardized dependency declarations.
Contract dependencies are often opaque, as they are typically hardcoded or dynamically resolved at runtime through deployed addresses. This poses several security risks.
 In this paper, we aim to study the width and depth of smart contract dependencies and the corresponding risks.

First, we devise an original methodology to study smart contract on-chain dependencies through historical interactions. 
Our sound methodology is based on querying a state-of-the-art of blockchain analytics platform.
We design and implement queries to answer four research questions on the scale, centralization, popularity, and transparency of smart contracts.

We run our queries on the Ethereum blockchain, up to the last block of December 31, 2024. Overall, we analyze over 41 million contracts and 11 billion interactions between contracts. We examine the results and their implications on security.
Our large-scale experiments yield the following findings.

\begin{enumerate}
    \item  RQ1: What is the scale of smart contract dependencies in the field? Most contract transactions involve multiple smart contracts (the median 2024 transaction triggers 4 distinct contracts). 
    This is clear evidence that there are risks associated with smart contract dependencies. 

    \item RQ2: Which are the most prolific creators of smart contracts? A handful of actors control a huge number of smart contracts (just 11 deployers are responsible for 50\% of all alive contracts). This is a highly centralized system with potential systemic risks, calling for awareness and mitigation.

    \item RQ3: Which smart contracts are most frequently called by other contracts? Over time, some smart contracts have become a cornerstone of the whole ecosystem. For those that are mutable, the major implication is that users' funds are at risk if the smart contracts' owners are compromised.
    
    \item RQ4: How transparent are smart contract interactions in major DeFi protocols? The actual interdependencies of DeFi contracts are way more complex than what is officially documented. This creates risks due to a fake sense of transparency. Our results call for better tools for documenting, tracking, and verifying smart contract on-chain dependencies.    
\end{enumerate}

To sum up, our contributions are:
\begin{itemize}
    \item A conceptual framework for reasoning about smart contract interactions and their risks. The framework is based on the contract call semantics of the Ethereum platform.
    \item A sound methodology to study smart contract interactions and dependencies based on on-chain analysis. We use the state-of-the-art blockchain analytics platform Allium.
    \item An empirical study of 41M smart contracts and 11 billion calls between smart contracts from the Ethereum blockchain, revealing major, original, and significant insights on smart contract usage in the field. We unveil significant risks caused by smart contract dependencies, calling for proper mitigation. 
\end{itemize}

\vspace{0.1cm}

\section{Background}
While numerous blockchain platforms exist~\cite{bhutta2021survey}, this paper focuses on Ethereum smart contracts.
Ethereum is the most widely adopted smart contract platform, powering a significant share of decentralized finance (DeFi), non-fungible tokens (NFTs), and other Web3 applications~\cite{werner2022sok,bao2022non}.
Its wide adoption and mature ecosystem make Ethereum an ideal target for studying smart contract interactions at scale.
This section introduces core concepts of Ethereum smart contracts, their deployment mechanisms, and their interaction via the Ethereum Virtual Machine (EVM).

\subsection{Smart Contracts}

A smart contract is a program deployed on Ethereum~\cite{wood2014ethereum}, identified by a unique address.
It consists of two main components: 1) Code defining the contract's behavior; 2) State representing on-chain data that can be updated via transactions.
Smart contracts may also hold a balance in ETH, Ethereum's native token.
Smart contracts are executed within the Ethereum Virtual Machine (EVM)~\cite{wood2014ethereum}, a Turing-complete virtual machine that defines the runtime environment for smart contracts.
The EVM operates deterministically: given the same inputs and blockchain state, it will always produce the same output.

\subsection{Smart Contract Deployment}
Smart contracts can be deployed in two primary ways:
\begin{enumerate}
    \item By Externally Owned Accounts (EOAs): EOAs are user-controlled accounts secured by private keys. They do not contain code. Developers typically deploy contracts by sending a special transaction from an EOA that includes the contract's bytecode. This transaction has no recipient and results in a new contract being assigned a unique address on the blockchain.
    
    \item By Other Contracts (Factory Pattern): Contracts can programmatically deploy other contracts using the \texttt{CREATE} or \texttt{CREATE2} opcodes. This allows for the dynamic deployment of multiple contract instances with shared logic but varying initialization parameters.
\end{enumerate}

The address of a newly created contract is deterministically derived from the creator’s address and a nonce (\texttt{CREATE}) or a salt value (\texttt{CREATE2}).
Once deployed, a contract’s code is immutable. However, its behavior can remain mutable based on external interactions. For example, a contract may be deployed with a function call provided by the user, and the contract’s behavior will depend on the address the user specifies.

Contracts can also be removed using the \texttt{SELFDESTRUCT} opcode. Historically, invoking \texttt{SELFDESTRUCT} would delete the contract’s code and storage from the blockchain and transfer its remaining ETH to a specified address.
However, following the Dencun fork in 2024~\cite{decun2024ethereum}, \texttt{SELFDESTRUCT} no longer removes a contract unless called in the same transaction as its creation.
In other cases, it simply transfers ETH, leaving the code and storage intact.

\subsection{Smart Contract Interactions}
\label{sc-interaction-bg}

Contracts interact with each other by:
\begin{itemize}
    \item Transferring ETH.
    \item Calling functions of other contracts.
\end{itemize}

The EVM supports several cross-contract interaction mechanisms~\cite{evmCodesOpcodes}:

\begin{itemize}
    \item \texttt{CALL}: Standard function invocation; maintains separate storage contexts.
    \item \texttt{STATICCALL}: Read-only call; cannot modify state.
    \item \texttt{DELEGATECALL}: Executes external code within the caller's context; useful for upgradeability, but introduces security risks. 
    \item \texttt{CALLCODE} (deprecated): Similar to \texttt{DELEGATECALL}, but rarely used in modern development.
\end{itemize}

These mechanisms enable the composition of complex, interdependent smart contracts — forming what is known as decentralized applications (DApps).

\subsection{Smart Contract Transparency}

Transparency is essential for building trust in smart contracts and DApps.

Unlike traditional software, smart contract source code must be made publicly available.
The best practice is that smart contract developers push their smart contract source code on platforms like Etherscan~\cite{etherscan} or  Sourcify~\cite{sourcifySourcifyeth}, a practice known as ``verified smart contract''.
Those platforms compile the provided source code to ensure that it matches the contract's bytecode on the blockchain. Then, they show the verification status in UI and API, giving users confidence that the deployed contract functions as the ones for which the source code is available~\cite{ethereumVerifyingSmart,ma2023abusing}.

To further enhance transparency, another best practice is that DApps publicly share their deployed contract addresses on their official websites or GitHub repositories.
However, this is done in an unstructured way, and there is no established standard or widely adopted method for documenting deployed DApp protocol addresses through machine-readable formats. 

\section{Experimental Methodology}
\subsection{Research Questions}

Our analysis of the smart contract ecosystem is guided by the following research questions (RQs):
\begin{itemize}

    \item \textbf{RQ1: What is the scale of calls between smart contracts in the field?}
    
    Understanding the scale of inter-contract interactions is crucial for assessing the complexity of the Ethereum smart contract ecosystem and identifying potential systemic risks. The interconnected nature of smart contracts has been overlooked by related work which studies contracts in isolation.
    
    \item \textbf{RQ2: Which are the most prolific creators of smart contracts?}
    
    Contract creation is a key driver of Ethereum's growth, but it is unclear whether it is widely distributed over many people or concentrated among a few dominant players. Identifying the most prolific contract deployers allows us to assess potential centralization risks.
    
    \item \textbf{RQ3: Which smart contracts are most frequently called by other contracts?}
    
    Frequently called contracts represent critical infrastructure within the Ethereum ecosystem. Identifying these highly used contracts highlights potential single points of failure. It helps prioritize security auditing efforts for components that could affect the broader system if compromised.
    
    \item \textbf{RQ4: How transparent are smart contract calls interactions in major DeFi protocols?}
    
    Transparency is a fundamental value proposition of blockchain technology, but the actual transparency of contract interactions may not live up to the expectations. Evaluating the gap between documented dependencies and on-chain reality helps users and stakeholders better to assess risks when interacting with decentralized applications.

\end{itemize}

\subsection*{\textbf{Data collection procedure}}
We answer our research questions through on-chain analysis.
Given the Ethereum mainnet's enormous size (2~TB on a full node and 2.7 billion transactions), we must employ a specialized blockchain database that efficiently stores blockchain data in order to extract the information we need. This is called a blockchain analytics platform.
To our knowledge, the three main ones are Allium, BigQuery, and Dune. 
We have compared them against ground truth data obtained from a self-hosted Erigon node.
Our results indicate that Allium delivers the highest accuracy. For example, Allium is the only platform that includes interaction data of precompiled contracts, which is required for our study. Besides, regarding the number of traces in 2024, the discrepancy between Allium and Erigon (ground truth) data is negligible, whereas BigQuery and Dune exhibited discrepancies higher than 7\%.
Consequently, in this paper, all numbers are extracted from Allium.
With various queries, we examine the entire Ethereum mainnet dataset from the genesis block through the final block of 2024, spanning over 21 million blocks. All queries are made available on a replication repository at \url{https://github.com/chains-project/crystal-clear}.

\subsection{Methodology}
\subsubsection*{\textbf{RQ1: Scale of Smart Interactions}}

We analyze the smart contracts and their interactions over multiple years of data from the Ethereum mainnet.
To address RQ1, we investigate the following aspects: 

\begin{itemize}  
    \item Number of contracts created (C), destructed (D), and alive (A).  
    Tracking contract creation and destruction trends provides insights into the lifecycle of smart contracts.
    The number of contracts that remain alive (i.e., created but not destructed) serves as a key indicator of ecosystem expansion and adoption.

    \item Distribution of contracts called per transaction:  
    Let $I_t$ be the number of unique smart contracts involved within a single transaction $t$.
    We analyze the distribution of $I_t$ for all transactions in a given time period.
    This metric provides insight into the degree of interconnectivity between contracts and the extent of their dependencies. To our knowledge, this aspect of smart contracts has never been studied before, as previous work tends to only consider smart contracts in isolation (e.g., \cite{huangSwordDamoclesUpgradeable2024}).
    
    \item Distribution of different types of calls per year: Let $P$ denote the total number of smart contract calls in a given year. Per the Ethereum semantics, these calls are of three distinct types: regular call $P_r$, delegate call $P_d$, and static call $P_s$. We analyze the distribution of $P_r$, $P_d$, and $P_s$ in $P$. This analysis provides valuable insights into the evolution of interaction activity and the specific dynamics of each call type, which is critical for understanding their role within the Ethereum ecosystem.
\end{itemize}

\subsubsection*{\textbf{RQ2: Contract Creation}}
To understand the dynamics of contract deployment, we analyze the most prolific contract creators in 2024. To identify them, we measure:  
\begin{itemize}  
    \item $C^{EOA}$: the number of contracts created by an EOA in a given timeframe. 
    \item $C^{contract}$: the number of contracts directly created by a given contract in a given timeframe. 
\end{itemize}  
We derive these metrics by filtering internal transactions that record the successful execution of the \texttt{CREATE} and \texttt{CREATE2} opcodes.

Contract creation plays a crucial role in shaping the smart contract ecosystem, as certain contracts act as factories, deploying numerous other contracts.
Identifying these prolific creators allows us to assess their influence on the network and detect potential centralization risks.

\subsubsection*{\textbf{RQ3: Most Called Contracts}} 
To determine the most widely interacted-with contracts, we analyze contract calls to identify the most called contracts on three non-deprecated call types, as mentioned in section~\ref{sc-interaction-bg}.

These metrics allow us to assess how broadly a contract is integrated into the ecosystem, beyond just raw call counts. Contracts with a high number of calls are likely to serve as a foundational component, widely relied upon across various DApps and applications.

This analysis also contributes to identifying critical points of centralization (focal contracts).
Centralization has potential reliability and security risks associated with a limited set of contracts.

\subsubsection*{\textbf{RQ4: Contract Transparency}}
Our goal is to examine the extent to which DApps  transparently document the smart contract interactions in their protocol.
By "transparent documentation," we refer to the practice of documenting all deployed addresses used by a DApp in its official protocol documentation. 
This is essential for users and stakeholders to assess the protocol security (verify audits) and manage their own usage (verify transactions before signing).

We analyze two major DeFi protocols: Uniswap, the most widely used decentralized exchange, and Lido, the most adopted liquid staking protocol~\cite{defillamaDefiLlama}.

We begin by manually collecting all contract addresses explicitly listed in the DApp’s official sources, such as official documentation websites and official GitHub repositories, in a reference set called $R$.
We then perform an on-chain analysis to identify all contracts interacted with by this initial set. 
This involves tracing historical calls made by the initial set and examining the target addresses.
Each discovered contract address is classified into one of three categories.
\emph{Documented addresses} are those officially listed by the protocol, from $R$.
\emph{Callback addresses} refer to external contracts not owned or deployed by the protocol, but used as callback in a transaction (e.g., token contracts or oracles).
\emph{Undocumented in-protocol addresses} are contracts not listed in any official source but appear to be part of the system, because being created by the protocol’s official deployers or being too tightly integrated into its logic that it cannot be considered as a callback.
To determine these types, we cross-reference documentation with on-chain data, review source code to identify external calls and analyze deployer addresses.
To our knowledge, this manual process is the only way to get a detailed view of how DApps transparently communicate their smart contracts.

\section{Experimental Results}

\subsection{RQ1: Scale of Smart Contract Interactions}
\begin{figure}[t]
\centering
\includegraphics[width=\linewidth]{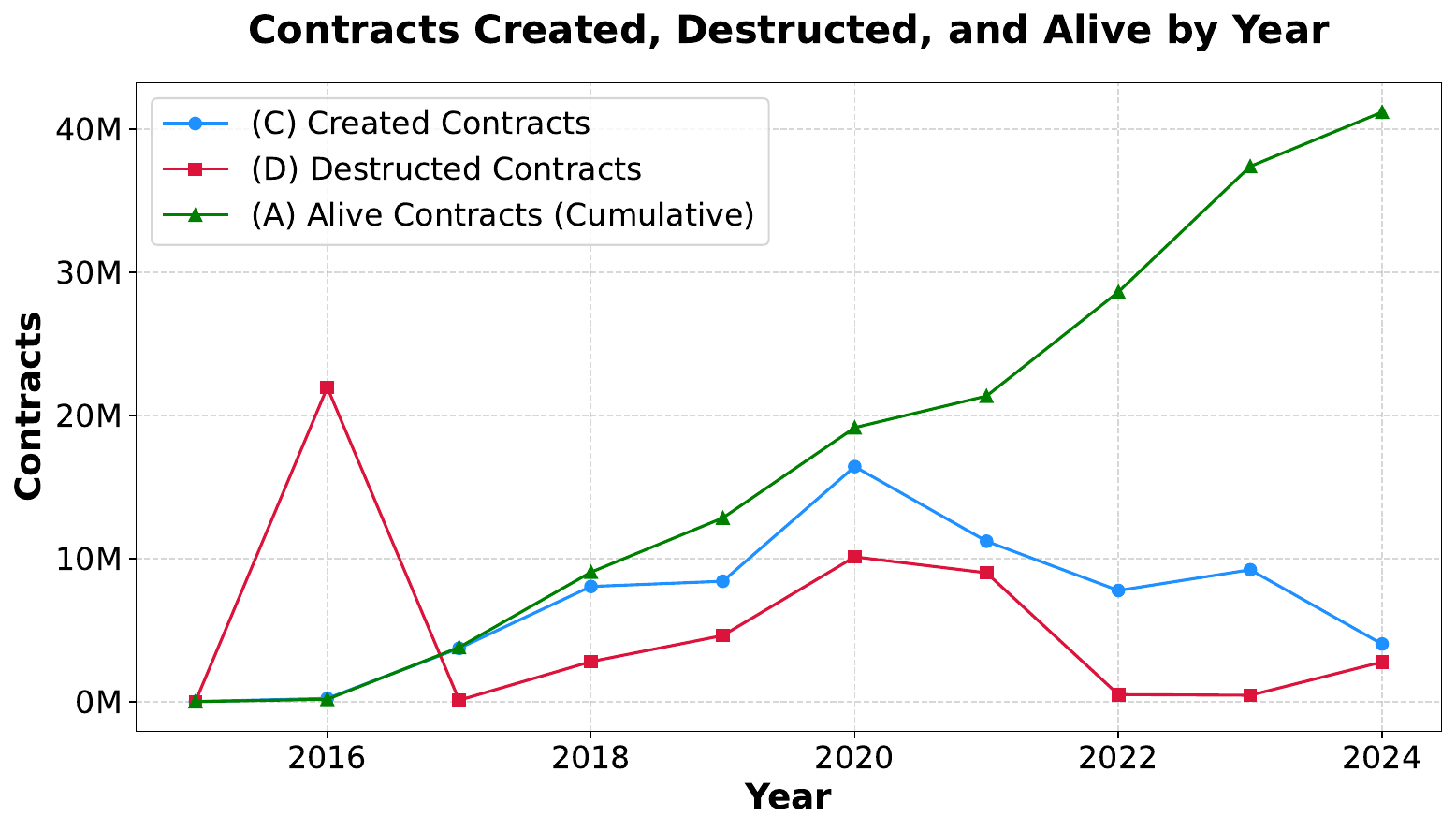}
\caption{RQ1: Number of created, destructed, and alive smart contracts (cumulative) per year on Ethereum mainnet. As of Dec 31 2024, Ethereum has 41M alive contracts.}
\label{fig:created_destructed}
\end{figure}

We start by looking at the true scale of the smart contract ecosystem on Ethereum.
While previous work has reported some figures, most of them are outdated by a few years at least~\cite{oliva2020exploratory}.

As of block number \href{https://etherscan.io/block/21525890}{21,525,890} (the final block of December 31 2024, 23h59), the Ethereum mainnet hosts $A = 41,199,792$ alive smart contracts (deployed, not destructed).

Figure~\ref{fig:created_destructed} illustrates the yearly trends in contract creation ($C$) and destruction ($D$), along with the total number of alive contracts ($A$) at the end of each year.
From Ethereum’s early years up until 2020, both $C$ and $D$ steadily increased, with 2020 being the most prolific year in terms of contract creation, corresponding to the DeFi summer~\cite{10.1093/jfr/fjad013}.
Regarding destruction, the peak observed in 2016 corresponds to a documented denial-of-service attack on Ethereum~\cite{ethereumEthereumNetwork}.


To assess contract interconnectivity, we analyze transactions involving smart contracts, excluding pure ETH transfers between EOAs.
Specifically, we measure the number of unique contracts involved in a single transaction ($I_t$).
When multiple contracts are called within the same transaction, it reveals a composition of contracts.
This metric helps us understand the extent of contract composition, a key aspect of Ethereum’s vision~\cite{ethereumSmartContract}.
Like Lego blocks, Ethereum smart contracts are designed to stack and interact, enabling modular and scalable applications.
Measuring composition shows how well real-world usage aligns with this principle.

\begin{figure}[t]
\centering
\includegraphics[width=\linewidth]{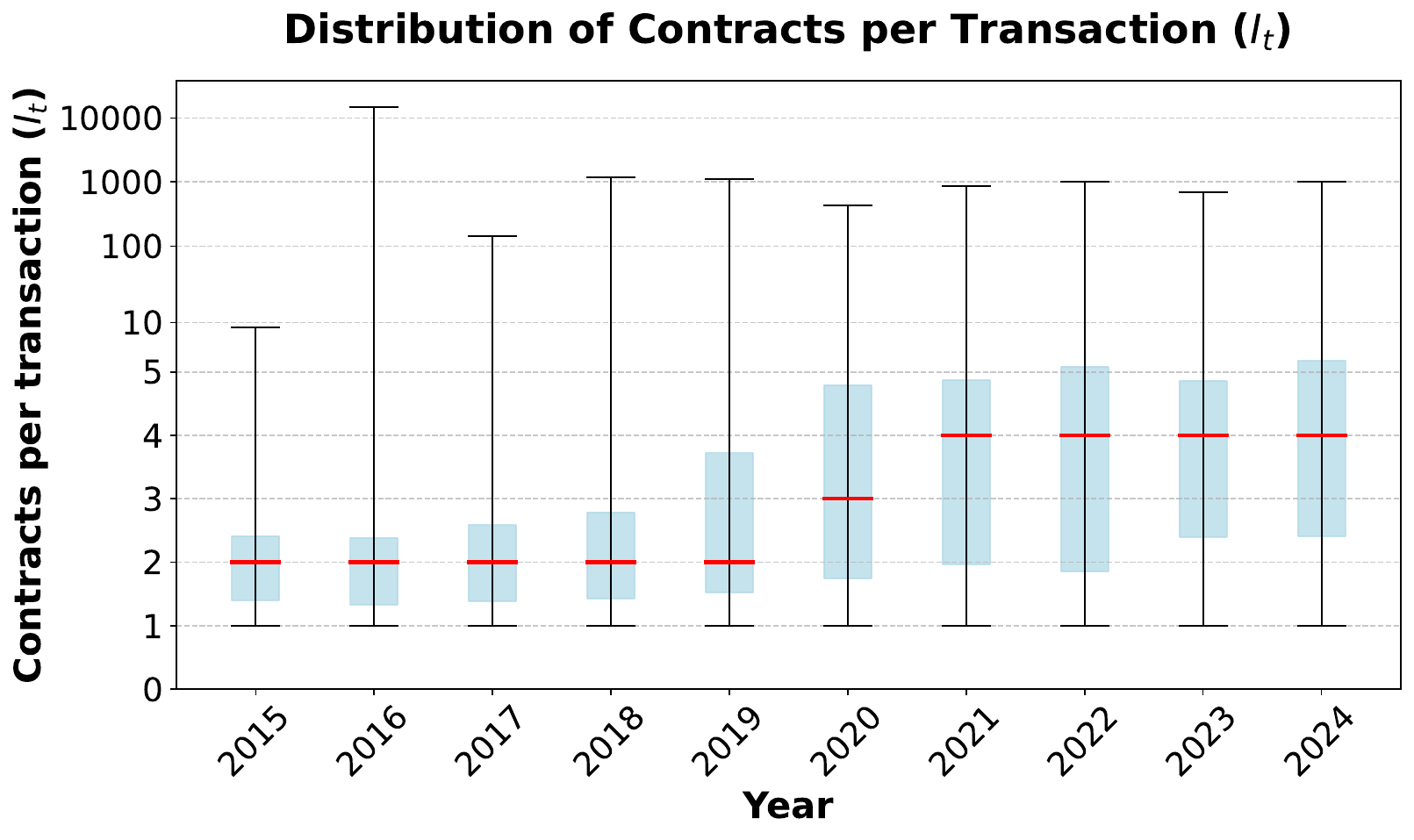}
\caption{RQ1: Distribution of contracts called per transaction per year on Ethereum mainnet (up to 2024). More than half of the transactions involve at least 4 contracts.
}
\label{fig:contracts_per_tx}
\end{figure}

Figure~\ref{fig:contracts_per_tx} presents the yearly distribution of $I_t$.
In 2024, the median number of unique contracts called per transaction is $I_t = 4$, with a maximum of $I_t = 1,004$ and a minimum of $I_t = 1$.  
The minimum value ($I_t = 1$) is consistent across all years, as every transaction must interact with at least one contract.
The maximum number of contracts per transaction has remained relatively stable.

Notably, the median $I_t$  keeps increasing from 2 (2015) to 4 (2024).
The same phenomenon happens for quartiles (the bottom and top ends of blue boxes).
The 25\% quartile has clearly increased in 2021 and 2023.
The 75\% quartile has seen a notable rise between 2019 and 2020, due to the DeFi summer, and again between 2023 and 2024.
This increase in median $I_t$ demonstrates ever greater contract reuse. Developers increasingly rely on composition based on enable modular decentralized application architecture.
Figure~\ref{fig:contracts_per_tx} is the evidence that the Ethereum ecosystem is moving to ever more permissionless composition, per the original vision \cite{wood2014ethereum}. A number of contracts are well designed, highly interoperable and efficient, allowing for the ecosystem to build on top of them.

\begin{figure}[t]
\centering
\includegraphics[width=0.8\linewidth]{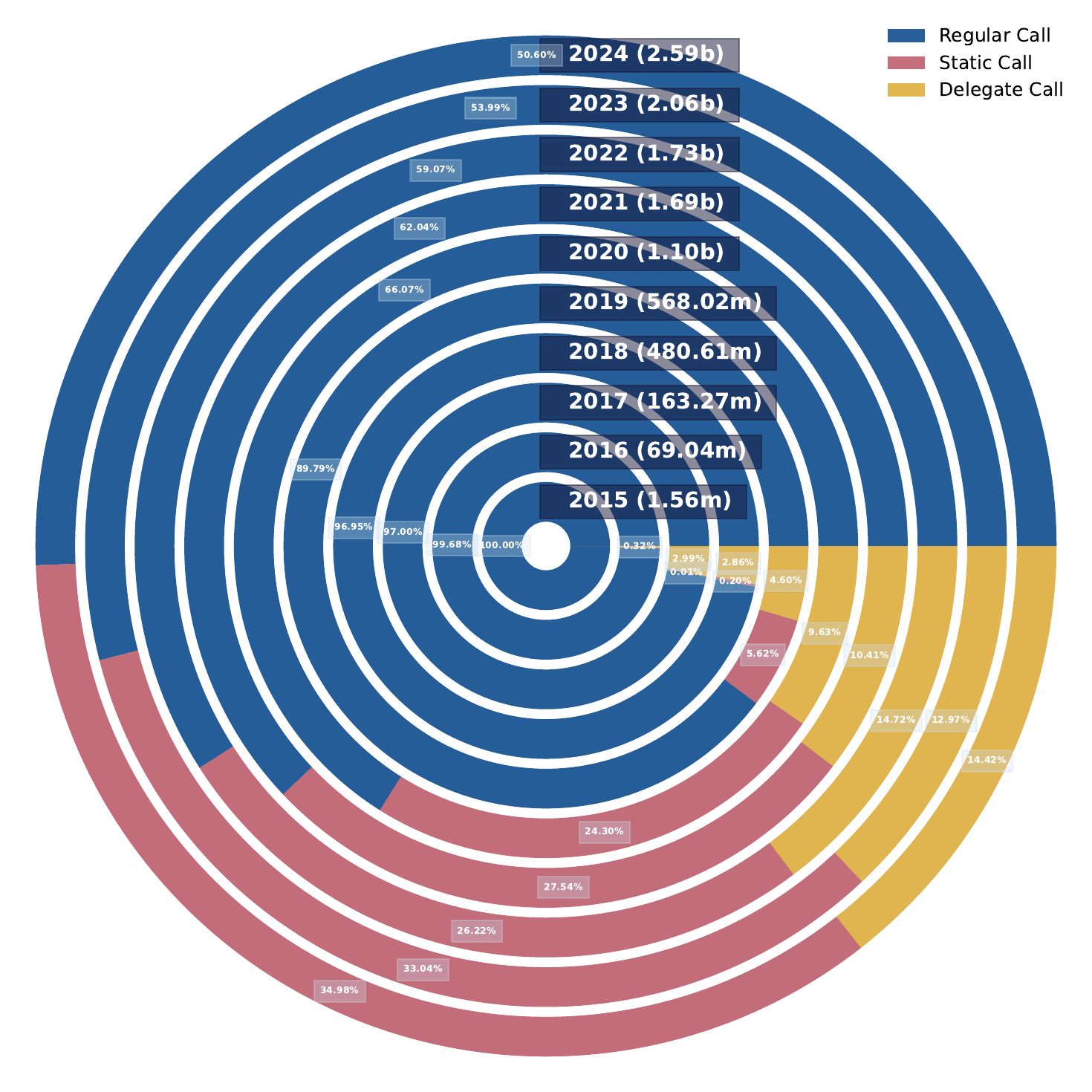}
\caption{RQ1: Evolution and distribution of each type of call (up to December 31 2024).
The significant increase of delegate calls indicate ever more usage of risk proxy pattern.
}
\label{fig:calls_yearly}
\end{figure}

To understand the evolution of contract interactions, we measured the number of successful calls (\(P\)) each year and computed the proportions of each call type, namely \( \frac{P_r}{P}, \, \frac{P_d}{P}, \, \frac{P_s}{P} \), corresponding to regular calls, delegate calls, and static calls, respectively. Figure~\ref{fig:calls_yearly} illustrates both the annual evolution of \(P\) and the distribution of call types. 
Over the past decade, the overall number of smart contract interactions has increased (from 1.56M to 2.59B), and also, the relative usage of call types has notably shifted. 
In the early years (2015–2017), regular calls dominated the interactions; however, starting in 2018, the usage of \texttt{DELEGATECALL} and \texttt{STATICCALL} began to increase, leading to a decline in the relative proportion of regular calls. By 2024, regular calls accounted for only 50\% of all interactions. 
It is worth noting that in 2020, there was a pronounced surge in the usage of \texttt{STATICCALL}, which might be attributed to the popular DeFi protocol contracts starting to use it.
Finally, the proportion of delegate calls also keeps increasing, up to 14.42\% in 2024.
This is explained by the popularity of the smart contract proxy pattern, and the need of dissociating data from behavior. Yet, it also poses security issues, which we will elaborate on that point in RQ3.

\begin{RQbox}{Answer to RQ1}
Our results clearly demonstrate the presence of smart contract composition on Ethereum. In 2024, the median number of smart contracts involved in a transaction was 4.
4\% of transactions involve more than 10 smart contracts and a few outliers involve up to thousands of smart contracts. To the best of our knowledge, our paper is the first to prove the reality of the vision of smart contract composition. This composition creates potential smart dependency risks, which is what we explore in the subsequent RQs.
\end{RQbox}

\subsection{RQ2: Contract Creation}

\begin{figure}[t]
\centering
\includegraphics[width=\linewidth]{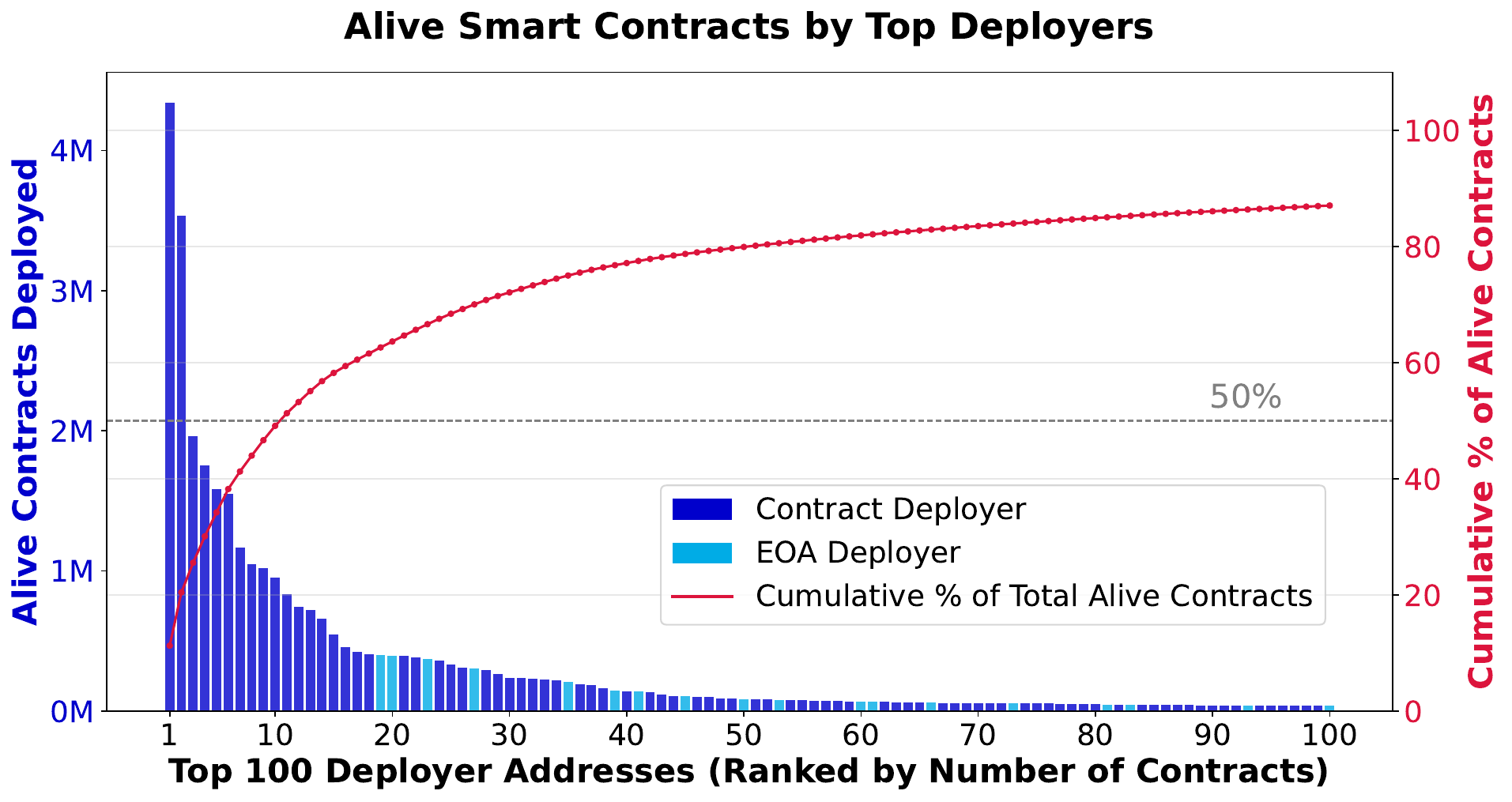}
\caption{RQ2: Top 100 deployers of contracts (alive in 2024) and their cumulative proportion. Eleven deployers are responsible for 50\% of all contracts.}
\label{fig:top100_deployers}
\end{figure}

We analyze the dynamics of contract creation and its relationship to contract dependencies.
With 41 million alive contracts on Ethereum, understanding who deploys them is essential.
Figure~\ref{fig:top100_deployers} illustrates the entities which deploy them, with the top 100 contract deployers.
Each deployer is represented as a bar, the height detailing the number of contracts  deployed (left y-axis) and the red line being the cumulative contribution to the total alive contracts (right y-axis).
First, we see that 83/100 top deployers are contracts themselves, with only 17 EOAs appearing in the top 100 list. This means that  Ethereum heavily uses the contract factory pattern.
The most prolific factory created 4.3M smart contracts (10\% of all alive contracts).
The top 11 contracts alone were responsible for deploying 50\% of all alive contracts in 2024, and collectively, the top 100 deployers accounted for over 80\% of all alive contracts.
This shows that the large number of contracts does not reflect a large number of smart contract developers and smart contract operators. There is actually a handful of entities responsible for the majority of contract creation. This confirms related work on centralization in Ethereum \cite{brown2023measuring}.

Table~\ref{tab:top10_deployers} provides a detailed overview of the top 10 deployers, including their human readable labels, source code availability (on Etherscan), activity status (indicated by at least one transaction in December 2024, whether the created contracts' behaviors are immutable (i.e., if a set of addresses has admin privilege to change the deployed contracts), the types of contracts they deploy, and the number of deployed contracts (same number as the corresponding bars on \autoref{fig:top100_deployers}.
For instance, Cointool BatchMinter~\cite{etherscanCoinToolBatch} has deployed 4.3M contracts, its code is verified, actively used and mutable.

It is positive to see that 6/10 contracts are verified, and surprising to see 4 unverified contracts responsible for creating 4.78M  child contracts.
Those contracts seem to be operated by crypto exchanges.
The \emph{Active} row clearly shows that some factories were heavily used in the past, yet have become dead since then, such as Bittrex (defunct company since December 2023)~\cite{bittrexglobalBittrexGlobal}.

Factory contracts, which deploy multiple instances from a single source, are attack vectors, as a vulnerability in them can spread to child contracts.
Table~\ref{tab:top10_deployers} shows that 8 out of these top 10 deployers primarily deploy proxy contracts, meaning that these child contracts interact with the implementation address using \texttt{DELEGATECALL}.
5 deployers implement EIP-1167~\cite{ethereumERC1167Minimal} proxies which means that all their created contracts are ``clones" and they redirect all their calls to one single address.
This has key implications for security.
If all child contracts are delegated to an address controlled by the deployer, this means that if the deployer is compromised, all child contracts become vulnerable as well, with their funds at risk. This is the case of Cointool BatchMinter and Bittrex Controller. For this reason, we strongly discourage this pattern.
If the child contracts are immutable and do not give special privileged access to the deployer, the fact that a single factory contract creates many child contracts has no immediate implication for control-based security.
This is the case for the 1inch CHI token and XENT token. However, even in such cases, immutability alone does not guarantee safety.
If the cloned logic itself contains vulnerabilities, all child contracts would inherit those flaws, making a single attack successful on all child contracts at once.
Thus, the widespread reuse of factory contracts amplifies the impact of vulnerabilities, regardless of upgradeability.

OpenSea’s now deprecated Wyvern protocol is a prime example of how proxy and factory design patterns can introduce centralization risks.
Under Wyvern, each user is assigned a personal proxy contract when they first interact with the marketplace.
These proxies are programmed to delegate their function calls to a centralized logic contract, a single implementation contract controlled by a master key at Opensea.
If this master key which controls the centralized implementation, were to be compromised, an attacker could manipulate all user proxies at once, putting their NFTs and assets under existential or ownership risk.

\definecolor{rowgray}{gray}{0.95}

\begin{table}
\centering
\scriptsize
\setlength{\tabcolsep}{1pt} 
\caption{RQ2: Top 10 deployer contracts responsible for 48\% of all alive contracts in 2024.}
\label{tab:top10_deployers}
\begin{tblr}{
  width = 0.85\columnwidth, 
  colspec = {
    Q[m,l]  
    Q[m,c]  
    Q[m,c]  
    Q[m,c]  
    Q[m,l]  
    Q[m,r]  
  },
  row{even} = {bg=rowgray},
  hline{1-2,12} = {-}{},
}
{\textbf{Deployer}}                  & {\textbf{Src.}\\\textbf{Code}} & {\textbf{Active}\\\textbf{Dec. 2024}} & {\textbf{Im-}\\\textbf{mutable}} & {\textbf{Deployed}\\\textbf{Type}}   & {\textbf{Deployed}\\\textbf{Contracts}} \\
{CoinTool:\\BatchMinter\\\href{https://etherscan.io/address/0x0de8bf93da2f7eecb3d9169422413a9bef4ef628}{0x0de..628}}   & \cmark     & \cmark    & \xmark      & {EIP-1167\\Proxy}     & 4,341,899 \\
{XEN Crypto:\\XENT Token\\\href{https://etherscan.io/address/0x0a252663dbcc0b073063d6420a40319e438cfa59}{0x0a2..a59}}      & \cmark     & \cmark    & \cmark      & {EIP-1167\\Proxy}     & 3,534,877 \\
{1inch:\\CHI Token\\\href{https://etherscan.io/address/0x0000000000004946c0e9f43f4dee607b0ef1fa1c}{0x000..a1c}}            & \cmark     & \cmark    & \cmark      & Gas Token          & 1,962,983 \\
{Coinbase:\\Commerce\\\href{https://etherscan.io/address/0x881d4032abe4188e2237efcd27ab435e81fc6bb1}{0x881..bb1}}          & \xmark     & \cmark    & --          & {EIP-1167\\Proxy}     & 1,753,886 \\
{Bittrex:\\Controller\\\href{https://etherscan.io/address/0xa3c1e324ca1ce40db73ed6026c4a177f099b5770}{0xa3c..770}}         & \cmark     & \xmark    & \xmark      & {Custom\\Proxy}       & 1,586,334 \\
{OpenSea:\\Registry\\\href{https://etherscan.io/address/0xa5409ec958c83c3f309868babaca7c86dcb077c1}{0xa54..7c1}}           & \cmark     & \cmark    & \xmark      & {Upgradeable\\Proxy}  & 1,548,230 \\
{Bitstamp:\\ Factory\\\href{https://etherscan.io/address/0xffa397285ce46fb78c588a9e993286aac68c37cd}{0xffa..7cd}} & \cmark     & \cmark    & \xmark      & {EIP-1167\\Proxy}     & 1,164,759 \\
{MMM BSC\\\href{https://etherscan.io/address/0x8a91c9a16cd62693649d80afa85a09dbbdcb8508}{0x8a9..508}}                     & \xmark     & \xmark    & --          & Unclear            & 1,049,380 \\
{Kraken:\\Deployer 2\\\href{https://etherscan.io/address/0xa24787320ede4cc19d800bf87b41ab9539c4da9d}{0xa24..a9d}}          & \xmark     & \cmark    & --          & {EIP-1167\\Proxy}     & 1,023,010 \\
{Contract\\Deployer\\\href{https://etherscan.io/address/0x46d781c076596e1836f62461f150f387ad140c0e}{0x46d..c0e}}           & \xmark     & \xmark    & --          & {Unknown\\Contract}       & 952,859  
\
\end{tblr}
\end{table}

Proxy risks are not theoretical. The recent Zoth exploit~\cite{mediumZothLose} confirms the real and immediate risks of centralized control in smart contract deployment.
The deployer wallet of the protocol had admin privileges to the proxy contracts. Once compromised, a malicious delegate was deployed, leading to the loss of \$8.4M.

\begin{RQbox}{Answer to RQ2}
The analysis reveals striking centralization in Ethereum's contract ecosystem, with just 11 deployers responsible for 50\% of all alive contracts and the top 100 
accounting for over 80\%. Most of these prolific deployers are contracts themselves rather than human-controlled accounts, highlighting the prevalence of the 
factory pattern in the ecosystem. This has particularly concerning security implications, as 8 of the top 10 deployers create proxy contracts that 
are significantly risky if the owner is compromised. Our original results challenge the perception of Ethereum as a highly distributed ecosystem, revealing 
instead a landscape where a handful of entities have disproportionate influence over the contract infrastructure, with varying degrees of transparency. Notably, 4 out of the top 10 deployers have unverified code with no available source code.
\end{RQbox}

\subsection{RQ3: Most Called Contracts}


\begin{figure}[t] \centering \includegraphics[width=\linewidth, trim=80pt 40pt 70pt 80pt, clip]{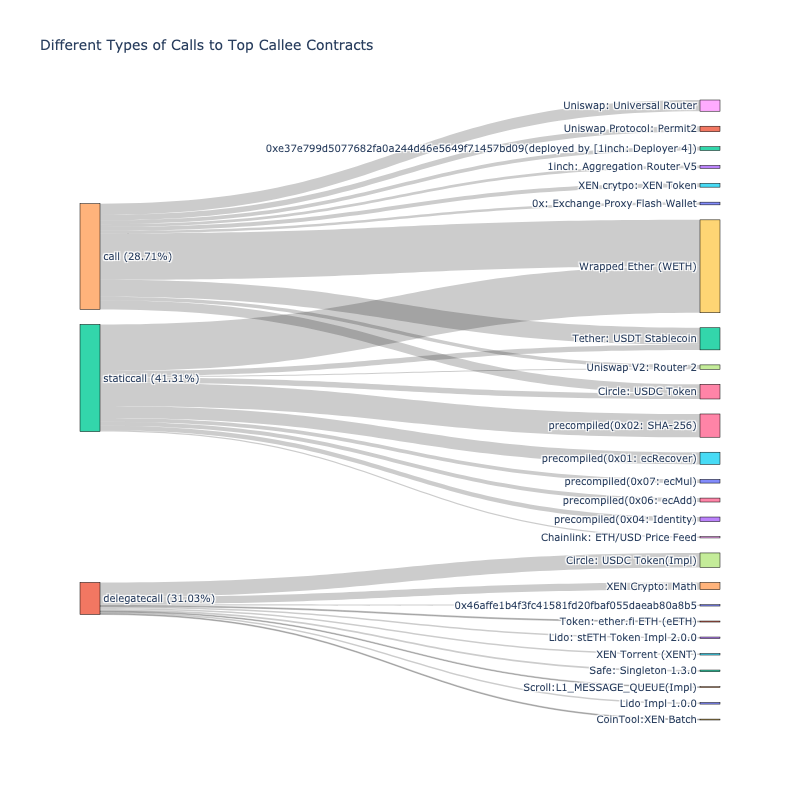} \caption{RQ3: Top callee contracts by call type on the Ethereum mainnet in 2024. Risk due to delegate calls should be carefully considered.} \label{fig:contracts_top_callee_2024} 
\end{figure}
On Ethereum, the most called contracts are those that serve as fundamental building blocks for other contracts.
We want to know what they are.
We consider the four call types in EVM, but \texttt{CALLCODE} was deprecated in November 2018 and is now negligible, so we exclude it in the analysis of RQ3.

Figure~\ref{fig:contracts_top_callee_2024} shows the result of most frequently called contracts.
First, we see that \texttt{CALL}, \texttt{STATICCALL}, and \texttt{DELEGATECALL} are all used in practice; they account for 28.71\%,  41.31\% and 31.03\% of the calls, respectively.
It means that the differences in their semantics have all found practical usage in successful DApp protocols.

We now look at \texttt{CALL}, the most standard call type, where the called contract maintains a separate state.
Figure~\ref{fig:contracts_top_callee_2024} shows the ten most called contracts with \texttt{CALL}.
First, for this type of call, we see that 4 out of the 10 top callee contracts are tokens (WETH, USDC, USDT, and Xen), confirming that tokens are fundamental building blocks of Ethereum, with \texttt{transfer} being the most frequently called function.

The remaining 6 out of 10 callee contracts are associated with decentralized finance. 
This empirically confirms that decentralized finance and programmable money a primary use cases of smart contract composition.
Because of the centrality of these contracts in the DeFi ecosystem, it is essential to pay special attention to their risk level. If one of them gets compromised, it may have contagion effects on the whole network of smart contracts.

Now we consider \texttt{STATICCALL}, which are pure, side-effect free calls, providing guarantees about transaction reversion if the call tries to change the state.
The WETH is heavily used in \texttt{balanceOf(address)}, \texttt{allowance(address, address)}, and \texttt{decimals()} for balance checking, authorization, and token handling.
Indeed, 3 of the top callee contracts are tokens (WETH, USDC, USDT), and their \texttt{STATICCALL} go in majority to \texttt{balanceOf}.
Interestingly, 5 out of 10 top callee contracts are precompiled contracts, which are in-protocol optimized for efficiency.
The fact that they are heavily used fully validate the design decision of Ethereum core team to have them in-protocol.

Out of the remaining two contracts called with \texttt{STATICCALL}:
1) the Uniswap Router contract used \texttt{WETH()} to return the address of the WETH, enabling accurate price retrieval.
2) the Chainlink Price Feed Contract is a blockchain oracle, a crucial component for obtaining external data for DApps. This distribution highlights the diversity of usages in \texttt{STATICCALL} interactions, confirming foundational infrastructures such as precompiled contracts and blockchain oracles for DeFi applications.

Finally, we look at  \texttt{DELEGATECALL}, which represents 31.03\% of the calls. \texttt{DELEGATECALL} enables to use the behavior of the callee contract on top of the state of the caller contract. It is essential for proxy contracts and upgrades, and also has other use cases.
5
out of the 10 top callee contracts with \texttt{DELEGATECALL} are related to tokens.
4 of them are implementation contracts of a proxy pattern (USDC, eETH, stETH, XEN Torrent), meaning that some tokens are implemented as proxies.
Notably, 3 of them have been upgraded in the past.
3 times for USDC, 4 for eEth, and 1 for stEth.
For example, the USDC contract has been upgraded 3 times, progressing from version 1.0 to version 2.0, followed by versions 2.1 and 2.2. 
This is a major security concern: if the token maintainers get compromised, it would be possible to steal the entirety of the token funds.
Overall, upgradable contracts are essential in the smart contract ecosystem of Ethereum, through \texttt{DELEGATECALL} interactions.
The flexibility of upgrading contracts comes with the risk that any upgrade to these contracts can significantly influence the overall system integrity, if not destroy the upgradable contract and its dependents.

\begin{RQbox}{Answer to RQ3}
Token contracts dominate the standard calls on Ethereum, with 4 of the top 10 called contracts being tokens, empirically confirming programmable money as the primary use case. Notably, among contracts using \texttt{DELEGATECALL}, 3 out of 4 token contracts have been upgraded in the past (e.g., USDC upgraded 3 times). This creates a systemic vulnerability: if token maintainers are compromised, entire token funds could be at risk. These findings demonstrate how certain contracts have become cornerstone infrastructure for the ecosystem, creating potential single points of failure with far-reaching implications for network security and stability.
\end{RQbox}

\subsection{RQ4. Contract Transparency}
To examine the extent to which DApps transparently document their protocol, we analyze two major decentralized exchange protocols: Uniswap~\cite{uniswapUniswapInterface} and Lido~\cite{lidoLidoLiquid}.

\begin{figure}[t]
\centering
\includegraphics[width=\linewidth]{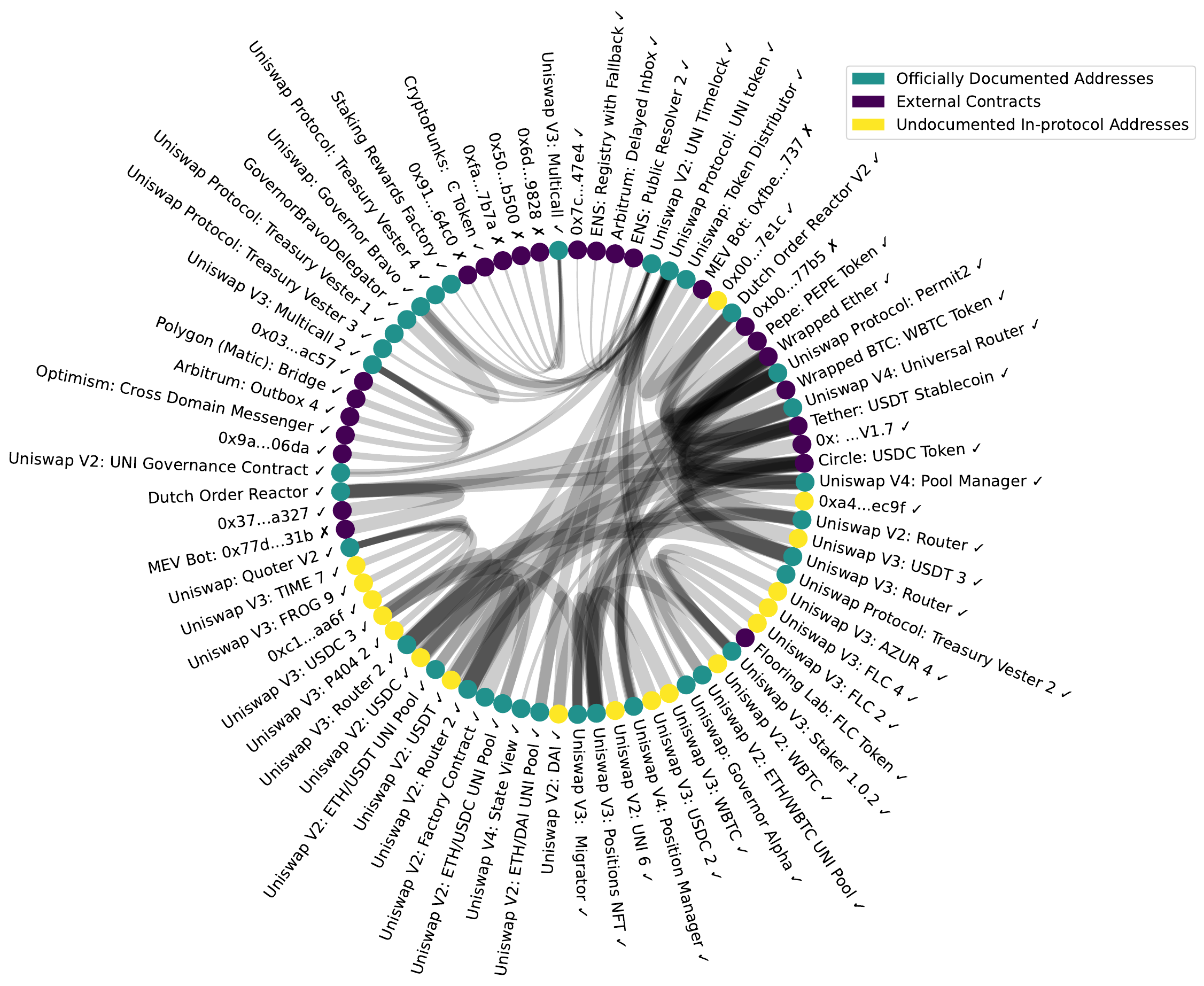}
\caption{RQ4 Case study: the smart contract interactions in the Uniswap protocol. Seventeen in-protocol contracts are not documented (yellow nodes).}
\label{fig:uniswap_network}
\end{figure}

Figures~\ref{fig:uniswap_network} and~\ref{fig:lido_network} illustrate each protocol's contract interactions:
\begin{itemize} \item \textbf{Green nodes} represent contracts that are officially documented by the protocol.
\item \textbf{Yellow nodes} correspond to undocumented but in-protocol contracts, deployed either by known protocol factories or by EOAs affiliated with the protocol.
\item \textbf{Purple nodes} indicate external contracts, such as user-supplied callback contracts or token addresses. These contracts are not a concern to transparency, as they are typically outside the scope of the primary contract logic.

\item \textbf{Edges} denote on-chain calls between contracts. Edge thickness reflects the frequency of calls, scaled logarithmically. \end{itemize}

\subsubsection{Uniswap case study}
We analyze the complete set of smart contracts officially documented by Uniswap, including contracts from all major protocol versions, UniswapX, and governance components such as timelocks and voting contracts.

Our goal is to examine the external interactions initiated by these contracts on-chain. Specifically, we identify and analyze all outbound calls made by documented Uniswap contracts. 

In total, we gathered 47 documented Uniswap addresses, of which 34 have interactions.
These contracts have interacted with 4,701,217 distinct external addresses.
To streamline our analysis, we examine the top 5 most frequently called target contract addresses for each documented contract, resulting in a set of 77 unique addresses, as illustrated in Figure~\ref{fig:uniswap_network}.

From the 43 distinct contracts that are not listed in the official documentation, we identify 18 in-protocol contracts created by Uniswap factory contracts or deployed by known Uniswap EOAs.
These include liquidity pools and utility contracts.
We claim that these contracts should be transparently documented in the official documentation.

We identify 25 out-of-protocol contracts, mostly callbacks, with no deployment link to Uniswap.
Of the 25 out-of-protocol contracts, 7 are well-known token contracts (e.g., USDC, USDT), which are expected in a decentralized exchange context.
The remaining 18 are callback contracts, addresses passed in by users during execution.

All official and in-protocol Uniswap contracts have verified source code, providing strong source code transparency and auditability.

Although Uniswap maintains a robust documentation portal, our analysis reveals that its token pool contracts remain undocumented despite being deployed, verified, and actively called by core components.
These contracts form part of the protocol’s functional surface area, and their addresses should be included in the documentation to improve transparency and support safe integration.

\subsubsection{Lido case study}

To confirm the findings from the Uniswap case study, we now look at another major DeFi protocol, Lido. Lido offers liquid ETH staking, allowing users to stake via tokens and earn staking rewards accordingly.

\begin{figure}[t]
\centering
\includegraphics[width=\linewidth, trim=40 20 20 20, clip]{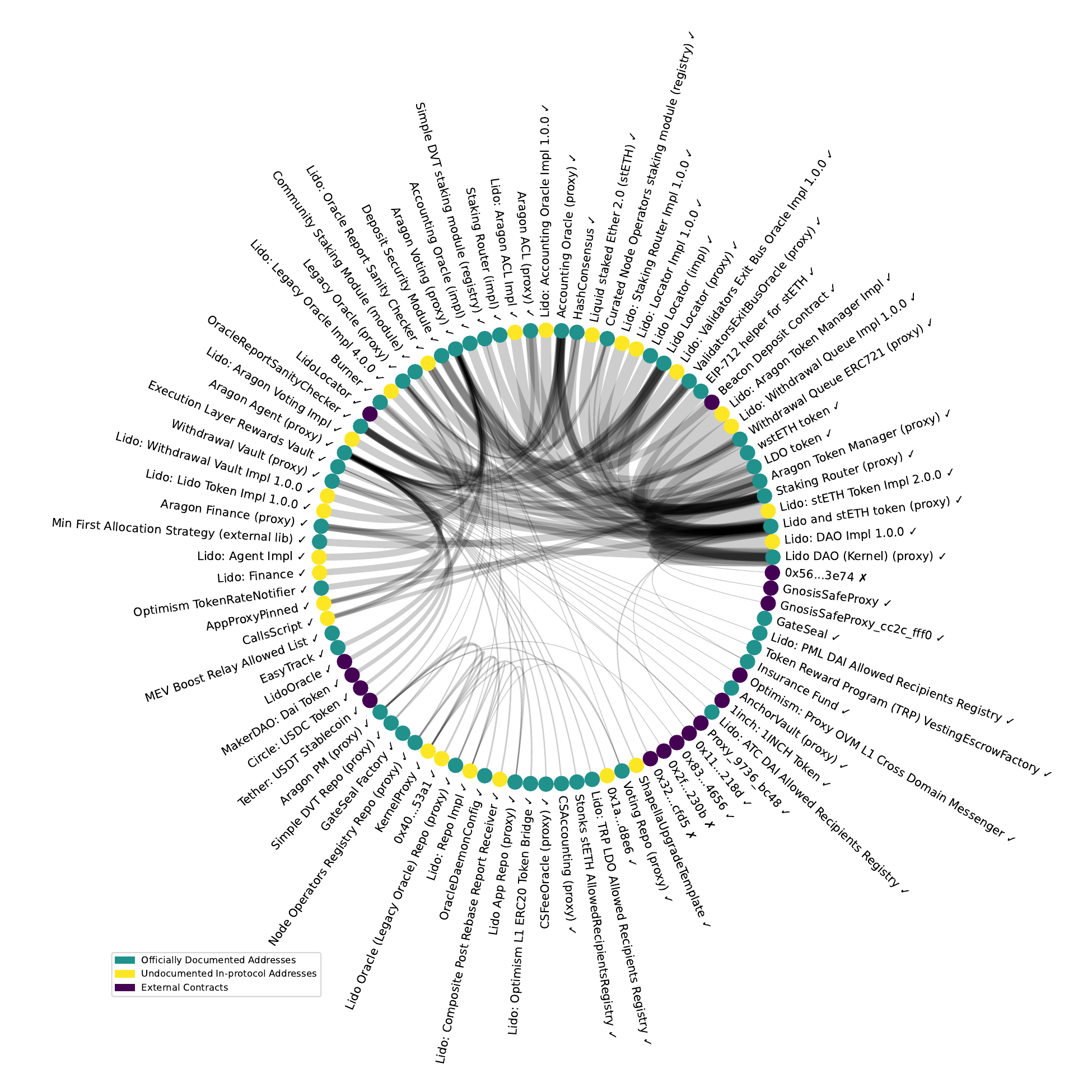}
\caption{RQ4 Case study: the smart contract interactions in the  Lido protocol. Lido does not transparently document all contracts in the protocol.}
\label{fig:lido_network}
\end{figure}

First, we extract 39 contracts from the official documentation~\cite{lidoMainnetLido} including Core Protocol, Oracle, and DAO contracts. Next,  we examined the calls following Block \#17244985, corresponding to the upgrade of Lido to its V2 version~\cite{lidoLidoVoting}.
We compute and visualize all interactions from those core contracts.
Figure~\ref{fig:lido_network} shows the result.
It reads like \autoref{fig:uniswap_network}.

Among the Lido contracts, some contracts are more interconnected than others, in particular Lido DAO, stETH, and the Staking Router, as can be seen on the right-hand side of Figure~\ref{fig:lido_network}.
Regarding transparency, 53 contracts in the network are included in the official documentation, whereas the other 41 are not documented. We manually analyze all of them: 16 are out-of-protocol contracts and 25 are by Lido-related deployer contracts or known EOAs.

Out of 25 contracts confirmed as Lido-related, 16 of them are implementation contracts, 4 are proxy contracts, 1 is a token contract, and 4 serve as functions such as OracleReportSanityChecker~\cite{etherscanLidoOracle}. This analysis demonstrates that the official Lido documentation is incomplete, with 25 contracts missing.
We argue that all contracts which play a key role in the protocol should be documented as such.





\begin{RQbox}{Answer to RQ4}
The analysis of Uniswap and Lido protocols reveals significant transparency gaps in DeFi documentation. Both ecosystems show concerning discrepancies between official documentation and on-chain data.
18 out of 52 and 25 out of 78 dependencies were not documented by Uniswap and Lido, respectively.
For protocol developers, this underscores the 
urgent need to improve deployment documentation practices.
For protocol users, it emphasizes the importance of understanding that official documentation may not 
capture the full scope of on-chain interactions and risks their funds are exposed to.
\end{RQbox}

\section{Threats to Validity}
\label{sec:discussion}

\emph{Internal Threats.}
The blockchain analytics platforms have data inconsistencies.
We tested BigQuery, Dune, and Allium and compared each platform's data against an Erigon node. Specifically, we assessed two metrics: the total number of traces recorded in 2024 and the trace count over a segment of 1001 blocks. For the total trace count, the results we have gotten from four data sources are different. Dune and BigQuery deviated by 7.2\% relative to the Erigon data, while Allium showed a negligible deviation of \(1.1 \times 10^{-5}\%\). Regarding the 1001-block segment, although Dune and BigQuery’s results matched those from the Erigon node, they omitted interactions with precompiled contracts; in contrast, Allium includes these interactions and the results are close to the results we have gotten from the Erigon node.

Additionally, the classification of interacted-with addresses for our case studies (Uniswap and Lido) was conducted manually. Authors cross-referenced historical transaction data and source code to label contracts.
While this process introduces potential for human error, it allows for fine-grained analysis that automated methods currently cannot provide.

\emph{External Threats.}
Our qualitative analysis of smart contract transparency is based on two case studies: Uniswap and Lido.
These findings may not generalize to all decentralized applications.
To mitigate this risk, we deliberately selected two of the most widely used and highly audited DeFi protocols.
Uniswap is recognized as the top decentralized exchange protocol, and Lido is the foremost liquid staking protocol on Ethereum~\cite{defillamaDefiLlama}.
Our analysis spans 171 contracts associated with these platforms, offering a robust foundation for understanding the transparency challenges faced by prominent DApps.

\section{Related Work}

\emph{Graph Analysis.}
Khan's survey \cite{khan2022graph} maps the related work on the graph analysis of Ethereum, up to 2021.
Early notable papers include  \cite{zhao2021temporal}
and \cite{liang2018evolutionary} which look at graph theoretic metrics such as in and out-degree in the Ethereum transfer graph. Lyu et al.~\cite{lyuAnalyzingEthereumSmart2022} used contract dependency graph with automated
analysis to analyze the impact of vulnerable contracts.
Beyond aggregated metrics, Ao et al. \cite{ao2022decentralized} made a deep dive on graph metrics for the Aave protocol.
Lee et al. \cite{lee2020measurements} study different interaction types on the Ethereum incl. contract-to-contract interactions.
None of this related work analyzes smart contract dependencies as we do in this paper.

\emph{Smart Contract Dependencies.}
A couple of tools have been developed to assist smart contract developers by providing graph visualizations of execution flows between smart contracts namely Smart-Graph~\cite{pierro2021smart} and MindTheDapp~\cite{10436108}.
They focus on analyzing individual DApps instead of the entire smart contract ecosystem within a blockchain (what this work does).
Aufiero et al.~\cite{aufieroDAppsEcosystemsMapping2024} used static analysis to study function calls within DApps and found that most DApp functionality is concentrated in a small subset of functions, particularly those in token contracts.
Several studies analyze on-chain data to map smart contract compositions.
Kitzler et al.~\cite{kitzlerDisentanglingDecentralizedFinance2023} used on-chain transaction data to conducted a topology analysis for DeFi protocols in the time span from January 2021 to August 2021.
Fröwis and Böhme~\cite{frowis2017code} measure the apparent trustless nature of smart contracts through a call graph of dependencies, finding out that two out of five smart contracts require trust in at least one-third party.
Ferreti and D'Angelo~\cite{ferretti2020ethereum} analyze the interactions between smart contracts through network modeling using historical transactions.
Serena et al.~\cite{serenaCryptocurrenciesActivityComplex2022a} analyzed the transaction record of four DLTs and specific characteristics of a network.
Harrigan et al.~\cite{harrigan2024token} study token compositions by constructing a graph of tokens with the events emitted by them.
To our knowledge, we are the first to provide a large-scale, up-to-date analysis of smart contract dependencies across the entire Ethereum blockchain, offering novel insights into the ecosystem's scale and transparency, as demonstrated in RQ1 and RQ4.

\emph{Smart Contract Upgradability.}
Huang et al.~\cite{huangSwordDamoclesUpgradeable2024} used a bytecode-based method to examine the prevalence of upgradeable smart contracts on Ethereum.
Qasse et al.~\cite{qasse2024immutable} examined the likeness of an upgradeable contract to be upgraded. Bodell III et al.~\cite{iiiProxyHuntingUnderstanding2023a} implemented a static analysis framework to analyzed proxy-based upgradeable smart contract. Li et al.~\cite{liCharacterizingEthereumUpgradable2024} used bytecode and transaction information to identify upgradeable contracts. Ruaro et al.~\cite{ruaro2024not} analyzed storage collision and identified potentially vulnerable contracts. None of them studied the fundamental risks of upgradable popular contracts, as we have done in RQ3.


\emph{Centralization.}
Collibus et al.~\cite{decollibusStructuralRoleSmart2022} analyzed the transaction networks of Ethereum and observed increasing 
centralization in the transaction networks. Juodis et al.~\cite{juodisExaminingTransactionalDecentralization2024} examined Ethereum transactions by analyzing metrics of transaction volumes. Lin et al.~\cite{linDefinitionDetectionCentralization2024} define and categorize centralization defects of smart contract code.
They do not study centralization through the lens of contract creation and ownership as we have done in RQ2.


\section{Conclusion}

Our comprehensive analysis of smart contract interactions on Ethereum reveals a complex ecosystem with significant implications for security, transparency, and 
decentralization. By examining 41 million contracts and 11 billion interactions, we have demonstrated that smart contract composition is not just theoretical but a 
practical reality. 

We have uncovered concerning centralization patterns, with just 11 deployers responsible for 50\% of all alive contracts, many implementing upgradeable proxy patterns that introduce systemic risks. Our analysis of the most called contracts 
identifies critical infrastructure components whose compromise could have cascading effects throughout the ecosystem. 

To sum up, our paper challenges the perception of Ethereum as a fully decentralized ecosystem, identifies key risks related to smart contract dependencies and call for better tools and practices for documenting, analyzing, and securing smart contract dependencies.

\section*{Acknowledgements}
This work was partially supported by 
the Swedish Foundation for Strategic Research (SSF);
the Ethereum Foundation;
the Swedish Governmental Agency for Innovation Systems (Vinnova).
We thank Christof Torres for the valuable feedback on the work.

\bibliographystyle{plain}
\raggedright
\bibliography{main}

\end{document}